\documentclass[pre,aps,amsmath,amssymb,reprint,floatfix,superscriptaddress]{revtex4-2}

\pdfoutput=1
\usepackage{xcolor}
\usepackage{siunitx}
\usepackage{graphicx}
\usepackage[colorlinks=true,citecolor=blue,urlcolor=blue]{hyperref}
\usepackage{orcidlink}
\usepackage{bm}

\def\av#1{\langle #1 \rangle}

\def\kB{k_{\rm B}}
\def\tf{t_{\rm f}}
\def\Ps{P_{\rm s}}
\def\Etot{E_{\rm tot}}
\def\e{{\rm e}}
\def\Teff{T_{\rm eff}}
\def\nev{n_{\rm ev}}

\newcommand{\LPENSL}{ \href{https://ror.org/02feahw73}{CNRS}, \href{https://ror.org/04zmssz18}{ENS de Lyon}, \href{https://ror.org/00w5ay796}{Laboratoire de Physique}, F-69342 Lyon, France}

\begin{document}

\title{Learning efficient erasure protocols for an underdamped memory}

\author{Nicolas Barros\,\orcidlink{0009-0000-1348-7725}}
\affiliation{\LPENSL}

\author{Stephen Whitelam\,\orcidlink{0000-0002-0086-6803}}
\email{swhitelam@lbl.gov}
\affiliation{Molecular Foundry, Lawrence Berkeley National Laboratory, 1 Cyclotron Road, Berkeley, CA 94720, USA}

\author{Sergio Ciliberto\,\orcidlink{0000-0002-4366-6094}}
\author{Ludovic Bellon\,\orcidlink{0000-0002-2499-8106}}
\email{ludovic.bellon@ens-lyon.fr}
\affiliation{\LPENSL}

\begin{abstract}
We apply evolutionary reinforcement learning to a simulation model in order to identify efficient time-dependent erasure protocols for a physical realization of a one-bit memory by an underdamped mechanical cantilever. We show that these protocols, when applied to the cantilever in the laboratory, are considerably more efficient than our best hand-designed protocols. The learned protocols allow reliable high-speed erasure by minimizing the heating of the memory during the operation. More generally, the combination of methods used here opens the door to the rational design of efficient protocols for a variety of physics applications.
\end{abstract}

\maketitle

\section{Introduction} To erase   one bit of information infinitely slowly requires the dissipation of $\kB T \ln 2$ of heat, where $\kB$ is Boltzmann's constant and $T$ the equilibrium temperature, a limit described by Landauer in 1961~\cite{landauer1961irreversibility,bennett1985fundamental}. Performing erasure at finite rate incurs additional heat costs that depend on how the bit is realized and which time-dependent protocol is used~\cite{Berut-2012,Jun-2014,Proesmans-2020,Dago-2021,Boyd-2022,Li-2024}. Experiments on a mechanical cantilever and phenomenological modeling indicate that a one-bit memory realized by an underdamped oscillator can be used to perform erasure more reliably and at higher speeds than can overdamped systems. However, two phenomena negatively affect the operation of underdamped memories: 1) residual damping losses are proportional to the velocity, and 2) the system heats up at low dissipation when heat transfer to the thermostat cannot compensate for the work done on the system. The latter energy overhead is described by an extended Landauer bound, where the temperature to consider is a weighted average of $T$ during the transformation~\cite{Dago-2022,Dago-2023-PNAS,Dago-2024-APR}.

Harnessing the potential of underdamped systems for logic operations~\cite{Ray-2021} requires finding optimal ways to perform erasure. Optimal protocols are not known because the problem is nonlinear~\cite{sanders2024optimal,sanders2024minwork} -- memories are bistable systems -- and we want to work in the limit of fast processes, where long-time approximations~\cite{blaber2021steps,ehrich2022energetic,rotskoff2015optimal} break down. In this article we show that evolutionary reinforcement learning~\cite{GA,floreano2008neuroevolution,such2017deep,whitelam2020learning}, a form of machine learning, can be used to find protocols for memory erasure in underdamped systems that are considerably more efficient than our best hand-designed protocols. To do so we express a time-dependent erasure protocol in the form of a deep neural network. Such an ansatz does not presuppose a functional form for the protocol, and is flexible enough to represent many different forms, including jumps and high-frequency features that appear in optimal protocols in other underdamped systems~\cite{gomez2008optimal}. Using a simulation model (or ``digital twin'') of the experiment, we apply a genetic algorithm to train the neural network in order to achieve a desired objective, which in this work is to render repeated erasures as reliable as possible, with a low energy of the memory after a single erasure as an intermediate objective. We then take the trained neural-network protocols and apply them to experiment, where we find that they behave as they do in simulation and represent a considerable improvement over our best hand-designed protocols, which were introduced in Refs.~\cite{Dago-2021,Dago-2022,Dago-2023-PNAS}.

In what follows we introduce the experimental realization of a 1-bit memory with a cantilever and our simulation model of the experiment. We show that the model is an accurate representation of experiment. We then use evolutionary reinforcement learning within simulations in order to derive efficient erasure protocols, and we apply these protocols to experiment, where we observe results consistent with simulation. The trained neural-network protocols allow reliable high-speed erasures in a physical model of a one-bit memory, potentially important for next-generation computing applications. A striking benefit of those protocols is that they are also low-energy: they keep the memory cool, increasing its reliability.  The combination of methods used here and the ability to deploy protocols in experiments opens the door to the rational design of hardware and software for deterministic $\kB T$-scale computing, and more generally to efficient protocols for a variety of physics applications.

\section{Experimental apparatus and simulation model}

\begin{figure*}[htbp]
	\centering
	\includegraphics[width=\linewidth]{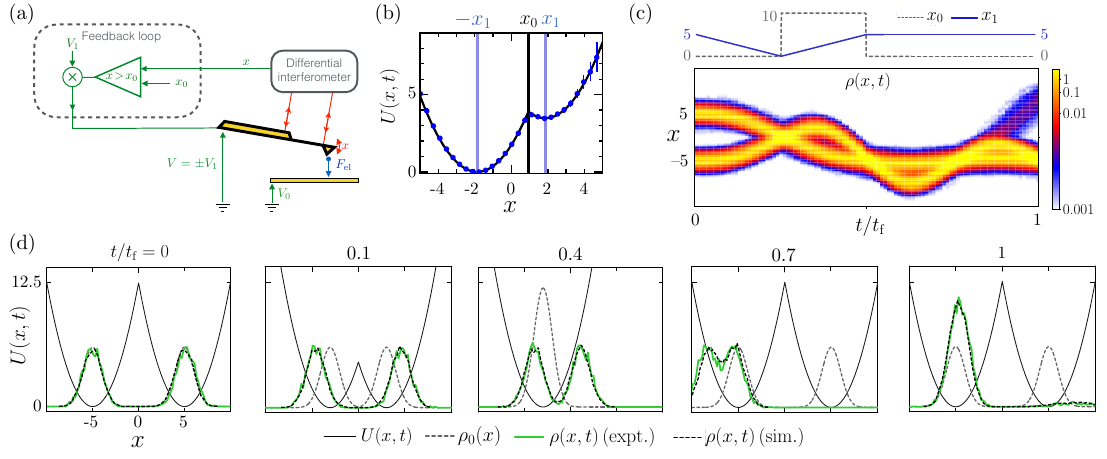} \vspace{-6mm}
	\caption{(a) Experimental setup: a conductive cantilever experiences an electrostatic force $F_\mathrm{el}$ due to the voltage difference $V-V_0$ with a facing electrode. The deflection $x$ of the cantilever, measured by interferometry, is compared to a threshold $x_0$ by the feedback loop setting $V=\pm V_1$. (b) This results in a tunable double-well potential $U$~\cite{Dago-2022-JStat,Dago-2024-Chapter}, that can be measured via the Boltzmann distribution $\rho_0(x)$ of the cantilever deflection in equilibrium: $U(x)=-\ln \rho_0(x)$. We plot this measured potential as dots, with error bars of one standard deviation, on top of the line denoting Eq.~\eqref{pot}. Logical states 0 and 1 are assigned to values $x<0$ and $x>0$, respectively. Our aim is to change the tunable parameters $x_0,x_1$ as a function of time in order to render multiple erasures (resets to 0 or 1) as reliable as possible. (c) Basic erasure protocol for a reset to state 0 (top). This protocol results in the distribution $\rho(x,t)$ in experiment (bottom). (d) Time-ordered snapshots with the potential $U(x,t)$, associated Boltzmann equilibrium distributions $\rho_0(x)$, and the distribution of cantilever positions $\rho(x,t)$ in experiment (green) and simulation (black dashed)~\cite{StatSingle}. The full movie of the time evolution of $U(x,t)$ and $\rho(x,t)$ corresponding to those snapshots is available as an ancillary file~\cite{SuppMatMovies}. Experiment and simulation are consistent. This basic protocol is only partially reliable: about $7\%$ of trajectories end in the wrong well. In what follows we apply machine learning to the simulation in order to learn more efficient erasure protocols $[x_0(t),x_1(t)]$, and we deploy these protocols in experiment.}
	\label{fig1}
\end{figure*}

Fig.~\ref{fig1}(a) shows our experimental setup. This consists of a micromechanical cantilever that operates, in the absence of external forces, as an underdamped harmonic oscillator characterized by a stiffness $k$, a resonance angular frequency $\omega_0 = 2\pi f_0 = 2\pi \times (\SI{1090}{Hz})$, and a quality factor $Q = 7$. The deflection $x$ of the resonator is measured using interferometry~\cite{Paolino-2013}. The oscillator is in thermal equilibrium with the surrounding air at room temperature $T$, and is subject to thermal fluctuations. The variance of $x$ at equilibrium is given by $\sigma^2 = \kB T/k \sim \SI{1}{nm^2}$. Subsequently, we express all lengths in units of $\sigma$, energies in units of $\kB T$, and time in units of $\omega_0^{-1}$. The fundamental oscillation period is thus $t_0=2\pi$.

Using a fast feedback loop~\cite{Dago-2022-JStat,Dago-2024-Chapter} we can modulate the electrostatic force acting on the cantilever so that it experiences a virtual energy potential $U(x,t)$ that is parameterized by two scalars, $x_0(t)$ and $x_1(t)$:
\begin{equation}
\begin{split}
U(x,t) = & \frac{1}{2} \big(x-S[x-x_0(t)]x_1(t)\big)^2 \\
 & + x_0 (t) x_1 (t)\big[S(x-x_0 (t)] +S[x_0 (t)]\big).
\end{split}
\label{pot}
\end{equation}
Here $S$ is the sign function, where $S[x] = - 1$ if $x < 0$, and $S[x] = 1$ otherwise. $U(x,t)$ has in general a double-well form, shown in Fig.~\ref{fig1}(b), where $x_0$ tunes the asymmetry and $x_1$ the barrier height. We start and end with a symmetric double well, and associate cantilever positions $x<0$ and $x>0$ with logical states 0 and 1, respectively. Starting in thermal equilibrium, with the cantilever in either well, we can impose a time-dependent protocol $[x_0(t),x_1(t)]$ in order to perform erasure, i.e. to bring the cantilever tip from its starting well to a specified final well in time $\tf =2 t_0$. We allow the protocol to act over one fundamental oscillation period $t_0$, and then wait in the final double-well for an additional time $t_0$ in order to assess the stability of the system. The time $\tf$ is very short, smaller than the relaxation time of the system $Q t_0/\pi$, and optimized protocols are required in order to minimize the kinetic energy of the cantilever post-erasure and so render erasure reliable. If the post-erasure kinetic energy is too large then the cantilever will escape from the intended well, rendering erasure unreliable~\cite{Dago-2024-APR}. In what follows our aim is to determine a protocol that makes multiple erasures, each of time $\tf$ and with no waiting time between them, as reliable as possible.

Our model of the underdamped oscillator consists of the dimensionless Langevin equation (see derivation in Appendix~\ref{langevin})
\begin{equation}
\label{lang}
\ddot{x}+ Q^{-1}\dot{x} = -U'(x,t)+\sqrt{2Q^{-1}} \, \xi.
\end{equation}
Here dots and primes denote differentiation with respect to time $t$ and position $x$, respectively, and $\xi$ is a Gaussian white noise satisfying $\av{\xi(t)}=0$ and $\av{\xi(t)\xi(t')}=\delta(t-t')$.

Introducing the cantilever velocity $v=\dot{x}$ and integrating Eq.~\eqref{lang} over the short time interval $\Delta t$ yields the update equations (details in Appendix~\ref{langevin})
\begin{align}
x(t+\Delta t)& =x(t)+v(t) \Delta t \label{ex},\\
v(t+\Delta t)&= \alpha v(t)- \left(1-\alpha\right) Q U'[x(t)] + \sqrt{1-\alpha^2}\tilde \xi(t), \nonumber 
\end{align}
where $\alpha \equiv \e^{-\Delta t/Q}$, and $\tilde \xi(t)$ is randomly drawn from a centered normal distribution of unit variance. We numerically integrate Eqs.~\eqref{ex} from time $t=0$ to $t=\tf$ using a timestep $\Delta t =1.09 \times 10^{-4} t_0$, starting in equilibrium with the potential in its symmetric double-well form.

 \section{Hand-designed erasure protocols and validation of the simulation model} In Fig.~\ref{fig1}(c) we present our basic erasure protocol~\cite{Dago-2021}. As shown in Fig.~\ref{fig1}(c-d), this merges the two wells symmetrically, translates the resulting single well to the left or right (for erasure to state 0 or 1, respectively), and then reconstitutes the double-well potential. The time-ordered snapshots in panel (d) show the potential (black), the associated Boltzmann equilibrium distributions (grey dashed), and the instantaneous distribution of cantilever positions in experiment (green) and simulation (black dashed)~\cite{StatSingle}. Our simulation results are consistent with our experiments. The cantilever positions are in general far from thermal equilibrium. From inspection of the panels at times $t/\tf =0.7$ and 1, and from panel (c), it is clear that in several trajectories the cantilever has so much kinetic energy that it escapes from the desired well prior to the end of the quiescent period. As a result, the basic protocol achieves erasure with only about 93\% success rate (in both experiment and simulation). 
 
\begin{figure}[tbp]
\centering
\includegraphics[width=6.93cm]{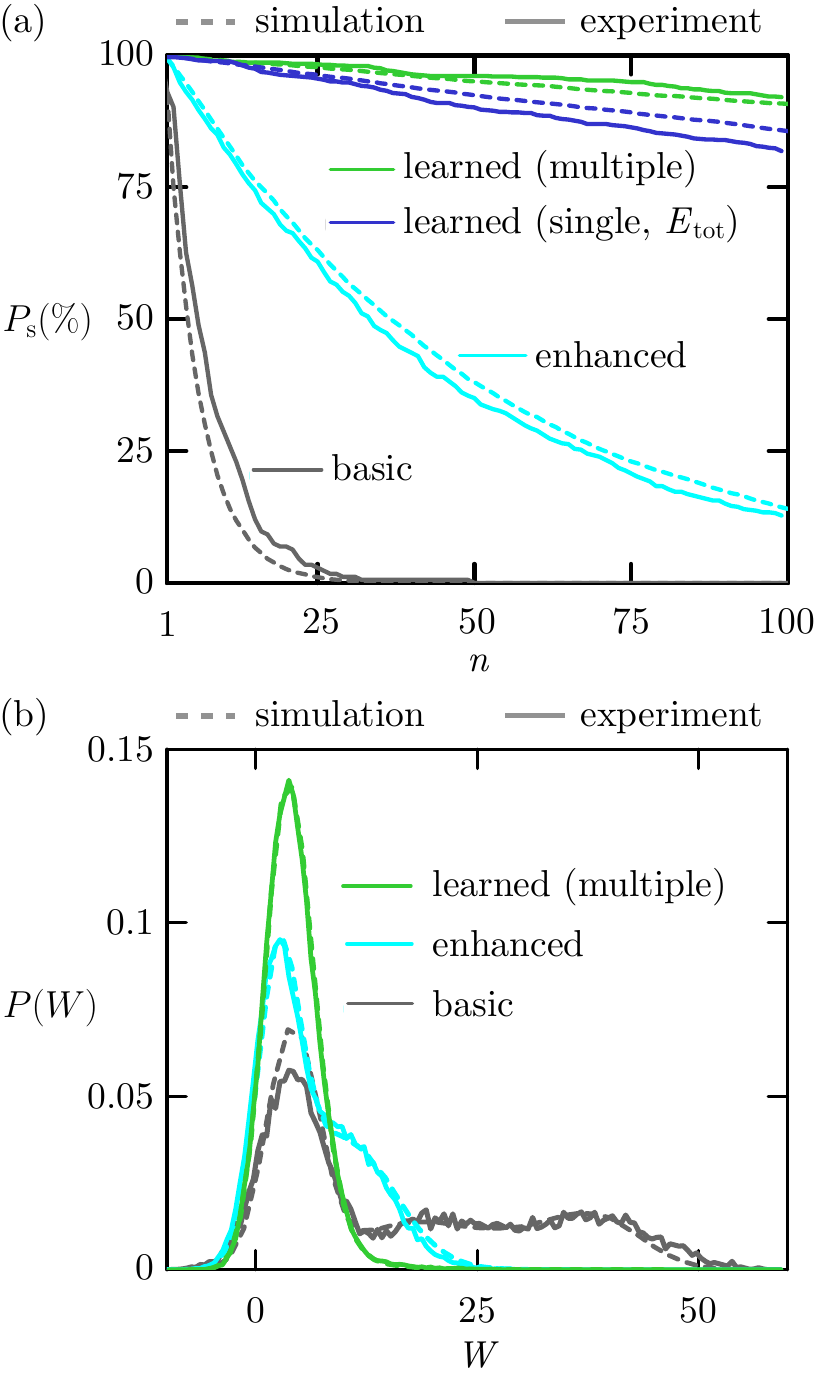}
\caption{(a) Success probability $\Ps$ as a function of the number of consecutive erasures $n$ for the four different protocols, and (b) distributions of work done per erasure calculated using 100 consecutive erasures, for three different protocols. We see agreement between simulation and experiment~\cite{ChainedStat}. The trained neural-network protocols are considerably more successful than our hand-designed protocols: $\Ps(100)=0$ for the basic protocol, $13\%$ for the enhanced protocol, $82\%$ for the protocol trained on a single erasure, and $92\%$ for the protocol trained on multiple erasures. The learned protocols also require less work than the hand-designed protocols.}
\label{figwdPs}
\end{figure}

\begin{figure}[tbp]
\centering
\includegraphics[width=\linewidth]{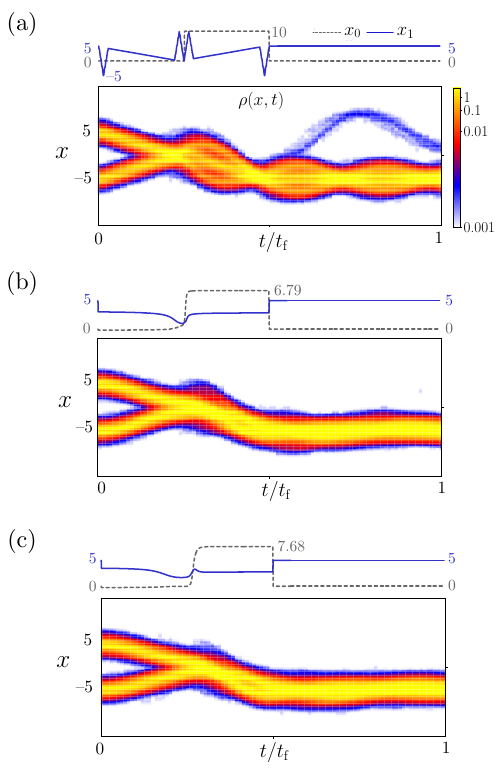}
\caption{Erasure protocols for a single reset to state 0 (top panels), and the resulting distributions $\rho(x,t)$ of cantilever positions in experiment. The format is the same as Fig.~\ref{fig1}(c). We show (a) the enhanced protocol, (b) the protocol trained to maximize the success rate and minimize the total energy upon performing a single erasure, and (c) the protocol trained to maximize the success rate of 100 consecutive erasures. In Fig.~\ref{figs4} in the appendix we show that distributions $\rho(x,t)$ from simulation and experiment are consistent. Movies of the time evolution of $U(x,t)$ and $\rho(x,t)$ for all protocols are available as ancillary files~\cite{SuppMatMovies}.}
\label{figprotocols}
\end{figure}

In Fig.~\ref{figwdPs}(a) we show that the basic protocol fails completely when applied to 100 consecutive erasures (with the target 0 or 1 chosen randomly with equal likelihood): the probability $\Ps(100)$ of observing 100 consecutive successful erasures is zero, in both simulation and experiment~\cite{ChainedStat}. The excess kinetic energy possessed by the cantilever following each erasure quickly builds and renders erasures completely unreliable. For all trajectories, we compute the work $W$ performed on the system during the erasure using the usual stochastic thermodynamics convention~\cite{sek10}, $W = \int \partial_t U(x,t) dt$. In the experiment, the feedback loop has no energetic contribution, because its only role is to construct the virtual potential~\cite{Dago-2022-JStat,Dago-2024-Chapter}. In Fig.~\ref{figwdPs}(b) we illustrate the work distribution associated with the basic erasure protocol in simulation and experiment, confirming again the accuracy of the simulation model, and underlining the fact that feedback used to create the virtual potential -- which is not considered in the simulation model -- has no energetic impact upon erasure. The work distribution is very broad, with a mean value $\av{W}=15.9$, far above the Landauer bound. With such a short $\tf$, the basic protocol is both unreliable and expensive.

In Fig.~\ref{figprotocols}(a) we show our best hand-designed protocol, an enhanced version of the basic protocol with delta function-like kicks at four points along the trajectory~\cite{Dago-2023-PNAS}. These kicks are motivated by the finding that optimal translation of an underdamped particle in a harmonic potential requires delta function impulses that abruptly change the velocity of the particle~\cite{gomez2008optimal}. In our case the enhanced protocol outperforms the basic protocol to a considerable degree, achieving a probability of success for one erasure of $\Ps(1) =99.2\%$. However, for 100 consecutive erasures the success rate drops sharply, to $\Ps(100)=13\%$ [Fig.~\ref{figwdPs}(a)]. The work distribution is much narrower than that of the basic protocol, with the average work decaying to $\av{W}=6.5$ [Fig.~\ref{figwdPs}(b)]. Part of the overhead to Landauer's bound originates from the kicks around $t/\tf=0.25$. These are sometimes in the wrong direction, creating the side lobes around the ideal trajectory in Fig.~\ref{figprotocols}(a). Some of these side-lobe trajectories eventually escape to the wrong well.

 \section{Learning optimized protocols}
 
 We can use the simulation model to produce an improved erasure protocol. Following~Ref.~\onlinecite{whitelam2023demon}, we encode the time-dependent protocol $[x_0(t),x_1(t)]$ in the form of a deep neural network, and train it using a genetic algorithm~\cite{GA,mitchell1998introduction} to achieve a desired goal (more details in Appendix~\ref{details}). This process is a form of evolutionary reinforcement learning. We have carefully benchmarked this procedure using overdamped passive-matter~\cite{whitelam2023demon,whitelam2023train} and active-matter systems~\cite{casert2024learning}, and in Appendix~\ref{benchmark} we show that it performs well in the case of an underdamped system whose optimal translation protocol is known exactly~\cite{gomez2008optimal}.

We instruct the learning algorithm to minimize the order parameter $\phi = 1-\Ps(1) + \av{\Etot}/100$ for single erasures. Here $\Ps(1)$ and $\av{\Etot}$ are the mean success rate for a single erasure and the mean total energy $\Etot = \frac{1}{2} v(\tf)^2+U(x(\tf))$ at the end of a single erasure (computed using $10^4$ independent trajectories). The value 100 is introduced in order to make the terms $1-\Ps(1)$ and $\av{\Etot}/100$ of similar order of magnitude (and therefore importance) for an erasure probability of about 99\%. Minimizing $\phi$ results in a protocol that performs erasure with high probability, and that results in a cantilever with small values of kinetic and potential energy. In Fig.~\ref{figprotocols}(b) we show the protocol produced by this procedure. It possesses jumps at the start and end, and rapid changes midway through. It performs well: its success rate for single simulated erasures is $99.9\%$. When applied to 100 consecutive erasures it produces a success rate of $85.6 \%$. Significantly, the protocol performs essentially as well in experiment: we measure success rates of $99.7\%$ for single erasures and $81.2\%$ for 100 consecutive erasures using the learned protocol with our experimental apparatus.

In Fig.~\ref{figprotocols}(c) we show the protocol produced by retraining the protocol of panel (b) to maximize the order parameter $\phi = \Ps(100)$, the success rate for 100 consecutive erasures (averaged over $10^4$ independent realizations of the stochastic process to ensure low statistical uncertainty). It is subtly different from the previous protocol, but performs considerably better on consecutive erasures: we measure success rates for 100 consecutive erasures of 90.9 \% (simulation) and 92.1 \% (experiment). That is, the learned protocol is essentially as effective when applied to 100 consecutive erasures as the basic protocol is for a single erasure. 

In Fig.~\ref{figwdPs}(a) we show the success rates $\Ps(n)$ for $n$ consecutive erasures for the two hand-designed protocols and the two learned protocols. We see agreement between simulation and experiment, with the two learned protocols outperforming the two hand-designed protocols. This is our central result: by applying machine learning to a simulation model of the experiment we have identified protocols that, when deployed in experiment, are considerably more reliable than our best hand-designed protocols. They are also cheaper: the average work during the 100 consecutive erasures is $\av{W}=4.2$ for the single trained protocol, and $\av{W}=4.1$ for the retrained one. The learned protocols are more narrowly distributed around their means than are the hand-designed protocols [Fig.~\ref{figwdPs}(b)]. We note that the learning algorithm was instructed to maximize the reliability of repeated erasures, not minimize work, but achieving reliability leads to less work expenditure than with the hand-designed protocols. 

The excellent agreement between the experiments and the numerical simulations on its digital twin show that the learned protocols are robust to non-idealities of the system, such as the description of the mechanical system and its frequency-dependent damping~\cite{Sader-1998,Bellon-2008} by a simple harmonic resonator. Using the dimensionless equations for the dynamics, we can in particular simply rescale the protocols for any variation of the resonant frequency $f_0$ or of the stiffness $k$. Small variations (of the order of $10\%$) of the quality factor $Q$ or final time $\tf$ have no noticeable effect on the learned protocols, but larger ones would require a retraining.

\section{Mechanism underpinning the success of the learned protocols}

\begin{figure}[tbp]
\centering
\includegraphics[width=6.6cm]{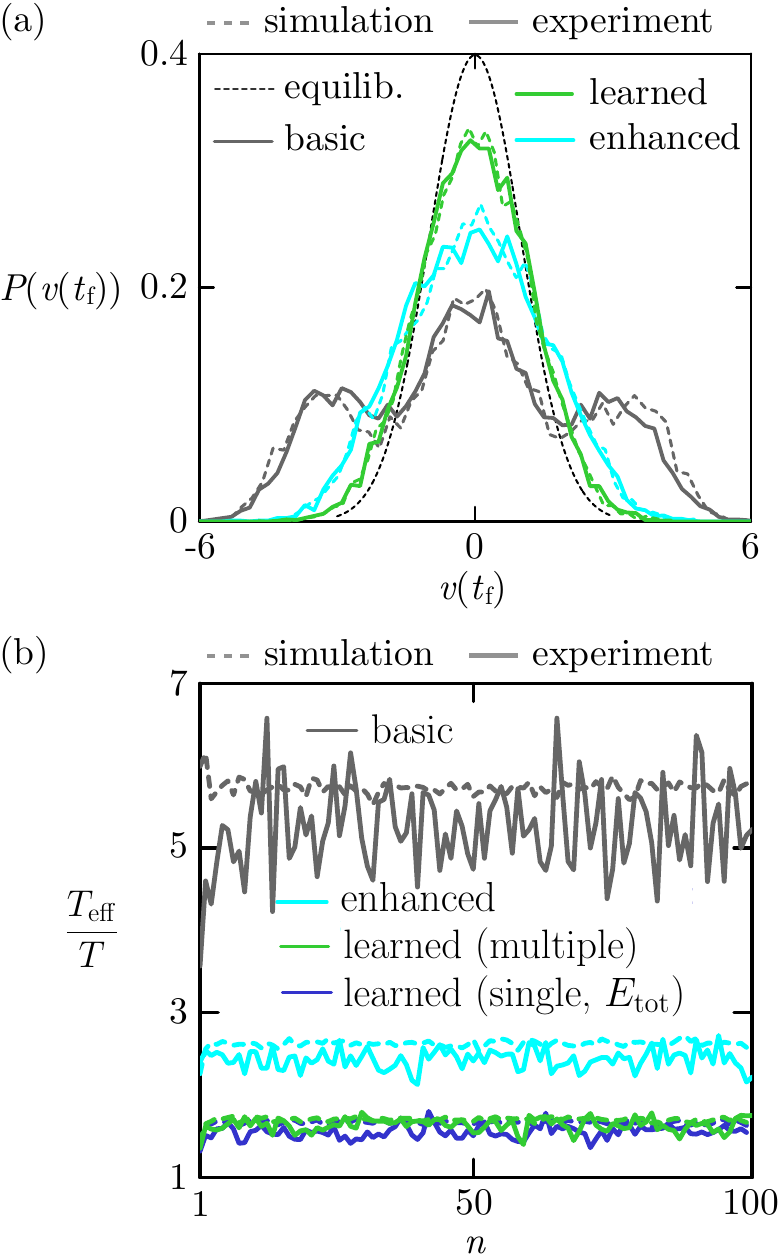}
\caption{(a) Distribution of cantilever velocities in simulation (dashed lines) and experiment (plain lines) for the basic, enhanced, and learned protocols, together with the initial equilibrium distribution (black dashed). Data are for single erasures in which the target is of state 0 or 1 with equal likelihood~\cite{StatSingle}. The learned protocol produces the narrowest distribution of velocities. (b) Effective temperature during repeated erasures after erasure $n$  in simulation (dashed lines) and experiment (plain lines), for the four protocols considered in this work~\cite{ChainedStat}. A major component of the success of the learned protocols lies in their ability to maintain a relatively low effective temperature.}
\label{fig_mechanism}
\end{figure}

The enhanced and learned protocols achieve a single erasure with high probability, 99.2\% and 99.9\%, respectively, and the distribution of cantilever positions at time $\tf$ is similar for the two protocols (see Fig.~\ref{figprotocols} and Fig.~\ref{figs3} in Appendix~\ref{data}).
The main difference between the two protocols is the kinetic energy of the cantilever after erasure. In Fig.~\ref{fig_mechanism}(a) we show distributions of cantilever velocities after one erasure for the basic protocol, enhanced protocol, and learned protocol of Fig.~\ref{figprotocols}(b), superposed with the equilibrium distribution at temperature $T$. The basic protocol has a long tail of high-energy values. The enhanced protocol is better, but the learned protocol is better still: the evolutionary optimization process was instructed to maximize erasure probability {\em and} minimize total energy (kinetic plus potential), and it has done this successfully. 

This difference in kinetic energy is a major component of the difference in consecutive-erasure success between learned and hand-designed protocols, shown in Fig.~\ref{figwdPs}(a). In Fig.~\ref{fig_mechanism}(b) we show the effective temperature, relative to thermal equilibrium, after $n$ consecutive erasures, $\Teff /T=\av{v(n \tf)^2}/\av{v(0)^2}$. The basic, enhanced, and learned protocols achieve steady-state effective temperatures of about 5.8, 2.6, and 1.6, respectively. With the resting barrier height of $\Delta E= 12.5$, the likelihood of spontaneous escape from the desired potential well during each erasure is $p_{\rm escape} \sim 2 \e^{-\Delta E/\Teff }$ (assuming that the basic attempt frequency is of order $f_0$, and so 2 attempts are made per erasure). The likelihood that no escape occurs in 100 erasures is then $\Ps(100) \sim (1-p_{\rm escape})^{100}$, which equates to 0, 0.19, and 0.92 if we use the values of $\Teff $ for the three protocols. These numbers are consistent with the measured values of 0, 0.13, and 0.92, and indicate that a major component of the success of the learned protocol lies in its ability to ensure a low kinetic energy or effective temperature during erasure.

The two learned protocols possess similar values of the effective temperature during repeated erasures, shown in Fig.~\ref{fig_mechanism}(b), but the protocol optimized for repeated erasure success does better on that metric than does the protocol optimized for a single erasure [Fig.~\ref{figwdPs}(a)]. This difference confirms that kinetic energy is not the only quantity that determines erasure success, and shows that the second learned protocol has adapted to ensure more reliable erasure in the face of this kinetic energy: the single-erasure protocol has evolved under exposure to initial conditions in equilibrium at temperature $T$, while the multiple-erasure training protocol has evolved under exposure to a steady-state initial temperature $\Teff >T$.

As illustrated in Fig.~\ref{fig_mechanism}(b), the memory reaches a non equilibrium steady state after only a few cycles. The choice of 100 consecutive erasures is thus more than sufficient to ensure that the resulting protocol can then be applied to a large number of consecutive erasures. The corresponding improvement in success rate over $N$ consecutive erasures done using the different protocols can be determined by extrapolating each line in Fig.~\ref{figwdPs}(a), assuming an exponential decay of $P_{\rm s}$ with $N$.

\section{Conclusions} We have used evolutionary reinforcement learning applied to a simulation model to identify efficient erasure protocols for an underdamped mechanical cantilever. We have shown that the same protocols when used in experiment are considerably more efficient than our best hand-designed protocols, essentially solving the heating problem that plagues repeated fast erasures. In this case the simulation model is an accurate representation of our experiment. If it were not, there exists the possibility of applying the learning algorithm directly to experiment, because all the information needed to train the neural network is accessible in experiment. This combination of methods therefore opens the door to the rational design of $\kB T$-scale hardware and their operating algorithms, and more generally to efficient protocols for applied physics.\\

The data supporting this study is available in~Ref.~\onlinecite{Barros-2024-Dataset}. Code for doing evolutionary optimization of cantilever protocols can be found at~Ref.~\onlinecite{sim_cantilever}.\\

\acknowledgments
This work has been partially funded by project ANR-22-CE42-0022. SW performed work as part of a user project at the Molecular Foundry at Lawrence Berkeley National Laboratory, supported by the Office of Basic Energy Sciences of the U.S. Department of Energy under Contract No. DE-AC02--05CH11231. SW was partially supported by the US Department of Energy, Office of Science, Office of Basic Energy Sciences Data, Artificial Intelligence and Machine Learning at DOE Scientific User Facilities program under Award Number 34532 (a digital twin for in silico spatiotemporally-resolved experiments).

\section*{Appendix}

\appendix

\section{Langevin equation non-dimensionalization and numerical integration}
\label{langevin}

The first oscillation mode of the cantilever is modeled by an harmonic oscillator of mass $m$, stiffness $k$, and damping coefficient $\gamma$. It is subject to a feedback electrostatic force $F_\mathrm{el}$ and thermal noise due to the contact with the thermostat. The dimensionful position $\bar x$ of the cantilever is described by the Langevin equation
\begin{equation} 
\label{eqlangevindim}
m \frac{d^2\bar x}{d\bar t^2} = -k \bar x - \gamma \frac{d\bar x}{d\bar t} + F_\mathrm{el} + \sqrt{2 k_B T \gamma} \xi,
\end{equation}
with $\bar t$ the dimensionful time. The electrostatic force can take only two values in our feedback implementation, which we set as $\pm k \bar x_1$ with a proper definition of the origin of $\bar x$. The natural scales of this problem are $\sigma=\sqrt{\kB T /k}$ for length, $\omega_0^{-1}=\sqrt{m/k}$ for time, and $\kB T$ for energy. We therefore introduce $\bar x = \sigma x$ and $\bar t = \omega_0^{-1} t$ into Eq.~\ref{eqlangevindim} in order to derive the dimensionless Eq.~\eqref{lang}, with the quality factor of the simple harmonic oscillator being $Q=m \omega_0/\gamma$.

To numerically integrate Eq.~\eqref{lang} we introduce the cantilever velocity $v=\dot x$, giving
\begin{equation}
\label{langv}
\dot{v}+ Q^{-1}v = -U'(x,t)+\sqrt{2Q^{-1}} \, \xi.
\end{equation}
Writing the left-hand side of \eqref{langv} as $\e^{-t/Q} \partial_t \left( v \e^{t/Q} \right)$, and assuming that $x$ is constant over the interval $[t,t+\Delta t]$, we multiply~\eqref{langv} by $\e^{t/Q}$, integrate over the interval $[t,t+\Delta t]$ and divide both sides by $\e^{(t+\Delta t)/Q}$ to obtain
\begin{equation}
\label{l1}
v(t+ \Delta t) -\alpha v(t) = -(1-\alpha) QU'(x,t) + \eta(t),
\end{equation}
where $\alpha \equiv \e^{-\Delta t /Q}$ and 
\begin{equation}
\eta(t) \equiv \e^{-(t + \Delta t)/Q} \sqrt{2 Q^{-1}}\int_t^{t+\Delta t} {\rm d}t' \, \xi(t') \e^{t'/Q}. 
\end{equation}
Since $\xi(t)$ is normally distributed so is $\eta(t)$, with correlations $\av{\eta(t)}=0$, $\av{\eta(t) \eta(t')} = 0$, and $\av{\eta(t)^2} = 1-\alpha^2$. Eq.~\eqref{l1} is the second line of Eq.~\eqref{ex}, with $\tilde \xi(t) = \eta(t)/\sqrt{1-\alpha^2}$ a random noise from a centered normal distribution of unit variance. The first line of Eq.~\eqref{ex} follows from the integration of $\dot{v} = x$ over $[t,t+\Delta t]$.

\section{Benchmarking the learning algorithm}
\label{benchmark}

\begin{figure*}[tbp]
 \centering
 \includegraphics[width=0.82\linewidth]{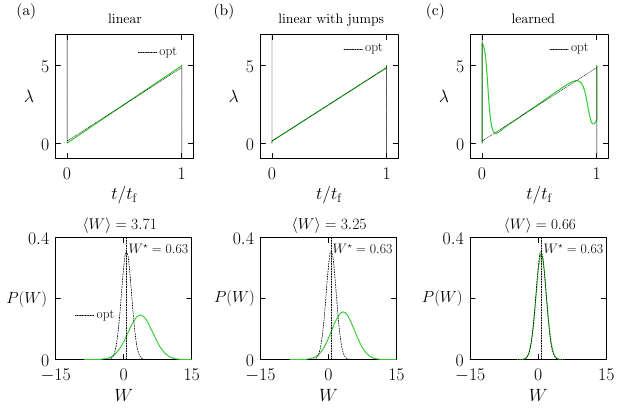} 
 \caption{Trap-translation problem of~Ref.~\onlinecite{gomez2008optimal} for an underdamped particle. We show trap position as a function of time (top) and work distributions (bottom) resulting from three protocols: (a) linear; (b) linear with jumps; and (c) a protocol expressed by a neural network trained by genetic algorithm to minimize $\av{W}$. The latter protocol is close to optimal. The black dashed line in the top panels is the optimal protocol \eqref{opt}. The black dashed vertical line in the lower panels is the optimal mean work value $W^\star$, Eq.~\eqref{opt_work}, and the dashed distribution is that resulting from the optimal protocol. Here averages and distributions are calculated from $5 \times 10^5$ independent trajectories. The initial velocity $v(0)$ and the initial position $x(0)$ are both chosen from a Gaussian distribution with zero mean and unit variance.}
 \label{figs1}
\end{figure*}

In this section we demonstrate the ability of the evolutionary learning procedure to identify efficient protocols for underdamped systems (we have previously demonstrated the same for overdamped passive- and active-matter systems~\cite{whitelam2023demon,whitelam2023train,casert2024learning}). 

We consider the trap-translation problem of~Ref.~\onlinecite{gomez2008optimal}, whose optimal solution is known exactly. An underdamped particle is subjected to an optical trap-like harmonic potential 
\begin{equation}
U(x,t) = \frac{1}{2}(x-\lambda(t))^2,
\end{equation}
where $\lambda(t)$ is the position of the trap center. The protocol that moves the trap from position $\lambda_{\rm i}=0$ to position $\lambda_{\rm f}$ in time $\tf$ with minimum work is given by
\begin{equation}
\label{opt}
\lambda^\star(t)=\Lambda (t+Q^{-1})+\Lambda \left[ \delta(t) - \delta(t-\tf)\right],
\end{equation}
where $\Lambda \equiv \lambda_{\rm f}/(\tf+2 Q^{-1})$. The work expended under this protocol is
\begin{equation}
\label{opt_work}
W^\star=\frac{\lambda_{\rm f}^2}{2+Q\tf}.
\end{equation}
We choose $\lambda_{\rm f} = 5$, $\tf=3.80$, and $Q=10$, in which case we have $W^\star \approx 0.63$.

To simulate the optimal protocol \eqref{opt} we must modify Eq.~\eqref{ex}: if we write $U'(x,t)$ as $U_0'(x,t) + \Lambda \left[ \delta(t) - \delta(t-\tf)\right]$, where $U_0'(x,t)$ is the force due to the first term in Eq.~\eqref{opt}, then the second line of Eq.~\eqref{ex} becomes 
\begin{equation}
v(t+ \Delta t)= {\rm r.h.s} + \Lambda \left( \delta_{t,0}-\delta_{t,\tf}\right).
\end{equation}
Here ``${\rm r.h.s}$'' denotes the right-hand side of Eq.~\eqref{ex} (with $U'(x,t)$ replaced by $U_0'(x,t)$), and $\delta_{i,j}$ is the Kronecker delta, equal to unity if $i=j$ and equal to zero otherwise. The work associated with the delta kicks is $\Delta W = \Lambda (v_{\rm i}-v_{\rm f})$, where $v_{\rm i}$ is the velocity immediately before the first kick and $v_{\rm f}$ is the velocity immediately after the second kick.

This problem poses a difficult task for any learning algorithm, because the optimal protocol contains delta functions. Moreover, the delta functions are important: in Fig.~\ref{figs1}(a) we show the results of a linear protocol $\lambda(t) = \lambda_{\rm f} t/\tf$, which produces mean work $\av{W} \approx 3.71$, and in Fig.~\ref{figs1}(b) we show the result of a linear protocol with jumps, given by Eq.~\eqref{opt} without the delta-function terms, which produces mean work $\av{W} \approx 3.25$. Both values of work are considerably in excess of the optimal value.

In Fig.~\ref{figs1}(c) we show the protocol expressed by a deep neural network trained by genetic algorithm to minimize $\av{W}$, following the procedure described in~Ref.~\onlinecite{whitelam2023demon}. This protocol has jumps at the start and end, and rapidly-varying features near the start and end. It produces mean work $\av{W} \approx 0.66$, only about 4\% larger than the optimal value. The distribution of work produced by the neural-network protocol is similar to that of the optimal distribution. 

The evolutionary learning algorithm has therefore identified a regularized version of the delta-function impulses seen in the optimal protocol, resulting in a protocol associated with a mean work only slightly larger than optimal. This comparison establishes confidence in the ability of the learning algorithm to identify efficient protocols for underdamped systems. In the main text we show that it can identify protocols for underdamped systems that are considerably more efficient than those designed by hand. The results of this section suggest that the rapidly-varying features seen in Fig.~\ref{figprotocols}(b-c) may become delta-function impulses in the true optimal limit, but also that the differences between the learned protocols and the hypothetical optimal ones are likely to be small.

\section{Training details}
\label{details}

We train the neural-network protocols following the procedure used in~Ref.~\onlinecite{whitelam2023demon}, specified by the code deposited in Ref.~\onlinecite{sim_cantilever}. Briefly, we simulate $10^4$ independent trajectories of the simulation model using a protocol $[x_0(t),x_1(t)]=[0,5]+{\bm g}_{\bm \theta}(t/\tf)$, where ${\bm g}$ is the two-dimensional output of a deep neural network whose input is $t/\tf$ and whose weights are ${\bm \theta}$. From these $10^4$ trajectories we construct an order parameter $\phi$ (described below). We then adjust the weights ${\bm \theta}$ using a genetic algorithm instructed to extremize $\phi$. To do so we run 50 lots of $10^4$ independent trajectories, each controlled by a neural network with slightly different randomly-chosen values of ${\bm \theta}$. We select the best 5 neural networks (the 5 associated with the best values of $\phi$), and clone and mutate these networks in order to produce a new set of 50 neural networks. Mutations consist of adding Gaussian random numbers $\epsilon \sim {\cal N}(0,10^{-4})$ to each weight of the neural network. We repeat this procedure until $\phi$ stops evolving, with $n_{\rm ev}$ used to indicate the number evolutionary steps taken.

\begin{figure}[tbp]
 \centering
 \includegraphics[width=0.85\linewidth]{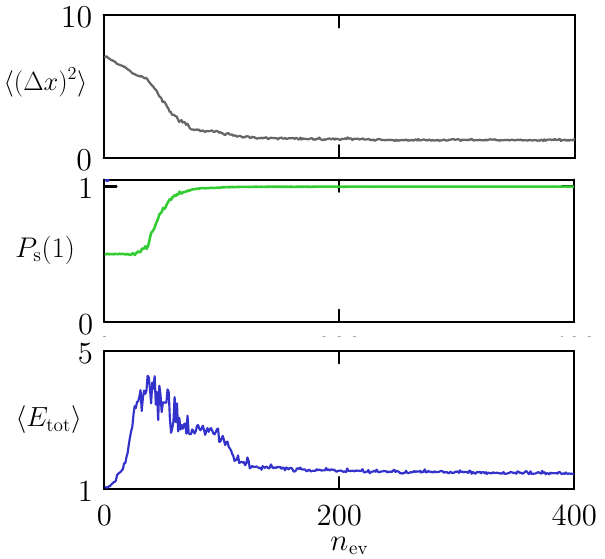} 
 \caption{Evolution of the components of the order parameter in Eq.~\eqref{phi1} as a function of the number $\nev$ of evolutionary steps, during the training of the neural-network protocol shown in Fig.~\ref{fig1}(c).}
 \label{figs2}
\end{figure}

To produce the protocol of Fig.~\ref{figprotocols}(b) we proceed as follows. We start from the null protocol, with ${\bm \theta}={\bm 0}$ and ${\bm g_{\bm 0}}={\bm 0}$, and so $x_0$ and $x_1$ are fixed to their initial values. We instruct the genetic algorithm to minimize the order parameter 
\begin{equation}
\label{phi1}
\phi =
\begin{cases}
 \av{(\Delta x)^2} + 50 & \text{if } \Ps(1) < 0.9; \\
 1 - \Ps(1) + \av{\Etot}/100 & \text{otherwise}.
\end{cases}
\end{equation}
Here $\av{(\Delta x)^2} =\av{(x(\tf)+5)^2}$ is the mean-squared distance between the final position of the cantilever and the final position ($-x_1(\tf)=-5$) of the center of the left-hand potential well, assuming erasure to state 0. For erasure to state 1, we use the same protocol with the sign of $x_0(t)$ reversed. $\Ps(1)$ is the mean success rate for a single erasure (erasure to state 0 is deemed successful if $x(\tf)<0$), and $\av{\Etot}$ is the mean total energy $\Etot = \frac{1}{2} v(\tf)^2+U(x(\tf),t_f)$ at the end of the protocol. The first clause of Eq.~\eqref{phi1} encourages the neural network to produce protocols in which the cantilever position ends close to the final position of the left-hand potential well. When the probability of erasure is greater than 90\%, the second clause becomes active and encourages protocols that 1) perform erasure with high fidelity, and 2) leave the cantilever with small values of total energy. 

Fig.~\ref{figs2} shows the evolution of $\av{(\Delta x)^2}$ (top), $\Ps(1)$ (middle), and $\av{\Etot}$ (bottom) as a function of the number $\nev$ of evolutionary steps. For each value of $\nev$ we show the values of these order parameters associated with the neural network that gives rise to the smallest value of $\phi$. The number of trajectories required after $\nev$ steps is $50 \times 10^4 \times \nev$. The trained protocol is then used in experiment.

The protocol of Fig.~\ref{figprotocols}(c) is obtained by taking the protocol of Fig.~\ref{figprotocols}(b) and retraining it using $10^4$ independent simulations, each of 100 consecutive erasures. In this case we instructed the learning algorithm to maximize $\phi=\Ps(100)$, the probability of obtaining 100 successful erasures. During the course of training we observed $\phi$ to increase from about 0.85 to about 0.9.

\section{Additional data}
\label{data}

\begin{figure}[!b]
 \centering
 \includegraphics[width=\linewidth]{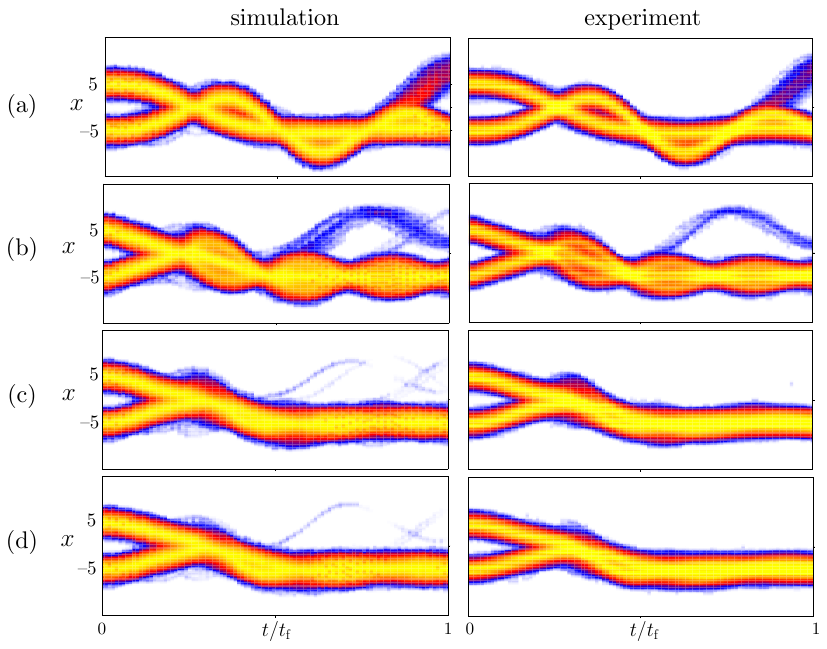} 
 \caption{Probability distributions $\rho(x,t)$ in the format of Fig.~\ref{fig1}(c) resulting from the four protocols considered in this work, in simulation (left) and experiment (right). Panel (a) shows the basic protocol of Fig.~\ref{fig1}(c). Panel (b) shows the enhanced protocol of Fig.~\ref{figprotocols}(a). Panel (c) shows the learned protocol of Fig.~\ref{figprotocols}(b), designed to maximize erasure fidelity and minimize total energy upon performing a single erasure. Panel (d) shows the learned protocol of Fig.~\ref{figprotocols}(c), designed to maximize the reliability of multiple successive erasures. In all cases we show a single erasure to state 0, starting from equilibrium~\cite{StatSingle}. Movies of the time evolution $\rho(x,t)$ for all protocols are available as ancillary files~\cite{SuppMatMovies}.}
 \label{figs4}
\end{figure}

\begin{figure*}[]
 \centering
 \includegraphics[width=\linewidth]{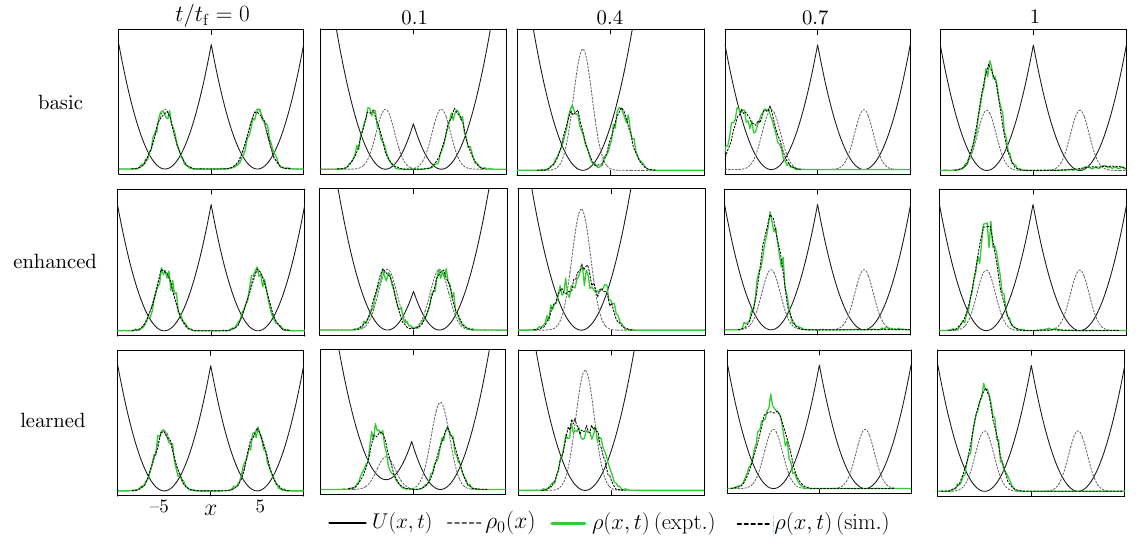} 
 \caption{Time-ordered snapshots for the basic and enhanced protocols, and for the learned protocol of Fig.~\ref{figprotocols}(c). We show the potential $U(x,t)$, the associated Boltzmann distributions $\rho_0(x)$, and the distribution of cantilever tip positions $\rho(x,t)$ in experiment (green) and simulation (black dashed) for a single erasure to state 0~\cite{StatSingle}. The enhanced and learned protocols achieve erasure with respective probabilities of 99.2\% and 99.9\%; the main difference between these protocols lies in the distribution of kinetic energies after erasure, as shown in Fig.~\ref{fig_mechanism}. Movies of the time evolution of $U(x,t)$ and $\rho(x,t)$ for all protocols are available as ancillary files~\cite{SuppMatMovies}.}
 \label{figs3}
\end{figure*}

In Fig.~\ref{figs4} we show the probability distributions $\rho(x,t)$ in the format of Fig.~\ref{fig1}(c) and Fig.~\ref{figprotocols} resulting from the four protocols. In Fig.~\ref{figs3} we show time-ordered snapshots in the format of Fig.~\ref{fig1}(d) for three protocols. In all cases, simulation and experiment are consistent. 

\bibliography{LearningErasureProtocols}

\begin{thebibliography}{37}
\bibitem [{Sta()}]{StatSingle}%
  \BibitemOpen
  \href@noop {} {}\bibinfo {note} {Results for single erasures are measured
  over $10^{4}$ independent trajectories in simulation, and $4 \times 10^{3}$
  independent trajectories in experiment.}\BibitemShut {Stop}%
\bibitem [{Sup()}]{SuppMatMovies}%
  \BibitemOpen
  \href@noop {} {}\bibinfo {note} {See videos included in the ancillary files
  of this arXiv preprint for movies showing the full time evolution of the
  potential $U(x,t)$ with the superposed distribution of $\rho(x,t)$ for the
  four protocols of this work.}\BibitemShut {Stop}%
\bibitem [{Cha()}]{ChainedStat}%
  \BibitemOpen
  \href@noop {} {}\bibinfo {note} {Results for 100 consecutive erasures are
  measured over $10^{4}$ trajectories in simulation. In experiment, we use
  $10^{3}$ independent trajectories for the trained neural-network protocols,
  $665$ independent trajectories for the enhanced protocol, and 174 independent
  trajectories for the basic protocol.}\BibitemShut {Stop}%
\end{thebibliography}

\end{document}